This article has been accepted by 24th Design, Automation and Test in Europe Conference (DATE) in 2021.# Dynamic Ternary Content-Addressable Memory Is Indeed Promising: Design and Benchmarking Using Nanoelectromechanical Relays

Hongtao Zhong, Shengjie Cao, Huazhong Yang, and Xueqing Li
BNRist, Department of Electronic Engineering
Tsinghua University, Beijing, China
Contact: xueqingli@tsinghua.edu.cn*Abstract*—Ternary content addressable memory (TCAM) has been a critical component in caches, routers, etc., in which density, speed, power efficiency, and reliability are the major design targets. There have been the conventional low-write-power but bulky SRAM-based TCAM design, and also denser but less reliable or higher-write-power TCAM designs using nonvolatile memory (NVM) devices. Meanwhile, some TCAM designs using dynamic memories have been also proposed. Although dynamic design TCAM is denser than CMOS SRAM TCAM and more reliable than NVM TCAM, the conventional row-by-row refresh operations land up with a bottleneck of interference with normal TCAM activities. Therefore, this paper proposes a custom low-power dynamic TCAM using nanoelectromechanical (NEM) relay devices utilizing one-shot refresh to solve the memory refresh problem. By harnessing the unique NEM relay characteristics with a proposed novel cell structure, the proposed TCAM occupies a small footprint of only 3 transistors (with two NEM relays integrated on the top through the back-end-of-line process), which significantly outperforms the density of 16-transistor SRAM-based TCAM. In addition, evaluations show that the proposed TCAM improves the write energy efficiency by 2.31x, 131x, and 13.5x over SRAM, RRAM, and FeFET TCAMs, respectively; The search energy-delay-product is improved by 12.7x, 1.30x, and 2.83x over SRAM, RRAM, and FeFET TCAMs, respectively.

*Keywords*—Ternary content addressable memory (TCAM), low-power, NEM relay, beyond-CMOS, dynamic memory.## I. INTRODUCTION

Various applications require frequent parallel data search to figure out whether the data of each row stored in a memory array match a given data stream [1][2]. These applications include routers, caches, database, etc. While such applications tend to be data-intensive in the big-data era, the demand on the searchable memory capacity is increasing. In addition, the memory update latency, search latency, reliability and power consumption can be also critical for both battery-powered gadgets and cooling-limited servers.

To achieve high parallelism, content addressable memory (CAM) has been widely used as it supports intrinsic *in-situ* data search without the need of pouring out data to the external for matching computing. Practically, ternary CAM (TCAM) is often adopted to support the extra '*don't care*' search rule claimed by the stored '*don't care*' bits or the input '*don't care*' bits. Fig. 1 shows the TCAM concept. Each cell compares the bitline inputs from the external in a differential XNOR style. The pre-charged matchline (ML) will be discharged by any mismatched cell on the same row. Thus, the settling-down behavior of the matchline in each row indicates the matching comparison result.

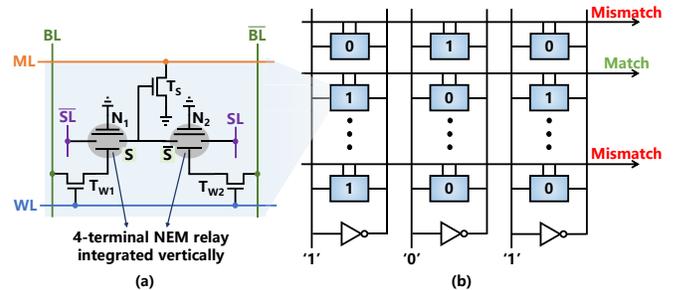

Fig. 1. Proposed 3T2N dynamic TCAM. (a) Cell; (b) Array search operation.

Existing TCAMs have been implemented using volatile CMOS and emerging nonvolatile memory (NVM) devices. Typical CMOS-based TCAM includes 16T SRAM structure [3] and 5T dynamic design [4]. The SRAM-based TCAM is mature and reliable, but needs 16 transistors for one bit. The CMOS dynamic TCAM (DTCAM) is denser but suffers from refresh problems in latency and energy. Nonvolatile TCAM can be achieved by STT-MRAM, RRAM, FeFET, etc. [5]-[8]. TCAMs using STT-MRAM or RRAM have higher write power and longer latency, and the low ON/OFF ratio along with device variation causes a leakage path and limits the achievable array size. FeFET, an emerging device with a high ON/OFF ratio [9]-[11], was used for dense lower-power TCAM. However, high-density FeFET TCAM suffers from disturbance problems, and current FeFETs are still in an early stage of low-voltage operation, resulting in long write latency (fast operation at high voltage causes endurance problems) [2].

To solve these problems, we explore the opportunities of using emerging devices towards dynamic TCAM with low refresh overheads and fast speed. Excitingly, this paper proposes the design of a novel dynamic TCAM, as shown in Fig. 1, with 3 transistors and 2 nanoelectromechanical (NEM) relays (3T2N). NEM relays are CMOS-compatible devices with moderate endurance, and have the unique characteristics of nearly zero OFF-state leakage, ultra-low ON-state resistance, and sharp ON-OFF transition with a hysteresis window [12]. In this paper, these NEM relay device features are harnessed to achieve fast search performance, low write power, moderate write latency, and a small footprint (the two NEM relays could be integrated on the top through the back-end-of-line (BEOL) process [13]).

Regarding the refresh issues of dynamic memories, our previous work has proposed One-Shot Refresh (OSR) to refresh all cells of an NEM relay memory array simultaneously with just one single write operation, without the need of prior row-by-row read operations [14]. With the custom circuit scheme in Fig. 1, this work also makes the refresh costs negligible by enabling the OSR scheme that refreshes the entire array with just one operation.

Itemized contributions include:

- The first NEM-relay-based dynamic TCAM to support fast and low-power search, low write power with moderate write latency, and low-cost refresh operations;
- Design and benchmarking of the proposed 3T2N TCAM, showing 2.31x, 131x, and 13.5x write energy efficiency over SRAM, RRAM, and FeFET TCAMs, respectively, and 12.7x, 1.30x, and 2.83x search energy-delay-product over SRAM, RRAM, and FeFET TCAMs, respectively.

In the rest of this paper, Section II reviews existing TCAM designs and the NEM relay background. Section III provides the details of the proposed 3T2N TCAM circuit design, including the circuit structure and operating methods. Section IV provides the circuit simulation results with analysis and discussions. Section V concludes this work.

## II. BACKGROUND

This section reviews existing TCAM designs using CMOS SRAM, RRAM, and FeFET. The introduction to NEM relay background is also provided to understand how the proposed TCAM works.

### A. Related Work about TCAM Designs

A well-known widely used TCAM design is the 16-transistor (16T) SRAM-based structure shown in Fig. 2(a). Besides, 5T CMOS-based dynamic TCAM design has been also proposed [4]. Recently, nonvolatile TCAMs are gaining more attention. An example TCAM design is shown in Fig. 2(b), with a 2-transistor-2-RRAM (2T2R) structure [6]. Compared with the SRAM-based TCAM, 2T2R TCAM saves area and the standby power but also induces other challenges. The first one is the higher write energy due to RRAM current-driven write mechanism and the longer write latency due to the conducting filament forming mechanism. Another challenge is the low ON/OFF resistance ratio worsened by the large device-to-device and cycle-to-cycle variations, which limits the yield of a large array size heavily.

Another promising TCAM solution based on FeFET has been proposed in [7] and [8], including the 4-transistor-2-FeFET (4T2F) structure and the 2-FeFET structure, as shown in Fig. 2(c-d). As FeFET is a capacitive load during write operation, its write energy can be significantly lower than RRAM. Besides, with an ultra-high ON/OFF ratio, it supports a larger array size. The 2-FeFET design is denser but is vulnerable to read and write disturbances [9]. Generally, FeFETs provide a promising approach to improved density over SRAM with moderate endurance. However, as an emerging device, continuous research is needed to ensure low-voltage operation, otherwise the write latency will be long or the write energy will stay high.

As mentioned above, prior design efforts of using dynamic memories, rather than the static memories (SRAM and NVM) suffers from the frequent refresh operations for conventional dynamic memories which interfere with normal TCAM search activities. The next sub-section will introduce the NEM relay device and also dynamic memories based on it. They will be used to build dynamic TCAM in this paper with no such data fresh problems.

### B. NEM Relay Characteristics

NEM relay is an emerging CMOS-compatible device and could be fabricated with 3 or 4 terminals [12]-[16]. Fig. 3(a) shows a 4-terminal (4T) NEM relay which consists of a drain electrode (D), a source electrode (S), a gate electrode (G) and a body electrode (B). A long bridge connecting the drain and source is controlled by the electrostatic force generated by the voltage difference between the gate and body ($V_{GB}$). When $V_{GB}$ exceeds a certain threshold, defined as the pull-in voltage ($V_{PI}$), the beam deflects towards the gate, which makes the bridge contact the drain and source. The device stays ON until $V_{GB}$ is below the pull-out voltage ($V_{PO}$), where the device returns to the OFF-state. Both $V_{PI}$ and $V_{PO}$ could be custom designed, and $V_{PO}$ could be much smaller than $V_{PI}$, leading to the $I_{DS}$–$V_{GB}$ hysteresis characteristic in Fig. 3(b). 4T NEM relay devices have a few intriguing features: (i) nearly zero OFF-state leakage due to the air isolation, and low ON-state resistance, leading to an ultra-high ON/OFF ratio; (ii) CMOS-compatible operation voltage and capacitive load during write operation, leading to ultra-low write energy; (iii) no threshold ($V_{th}$) drop as a pass transistor because of direct contact between the drain and the source. As to be revealed subsequently, these characteristics are exploited by the new TCAM design, leading to improved power and latency performance.

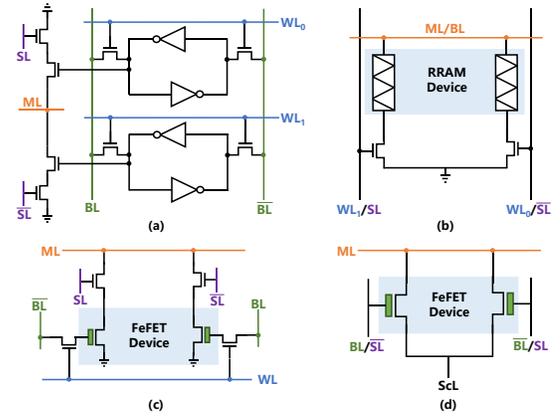

Fig. 2. Some typical TCAM designs. (a) 16T CMOS [3]; (b)2T-2R ReRAM [6]; (c)4T-2FeFET [7]; (d) 2 FeFET [8].

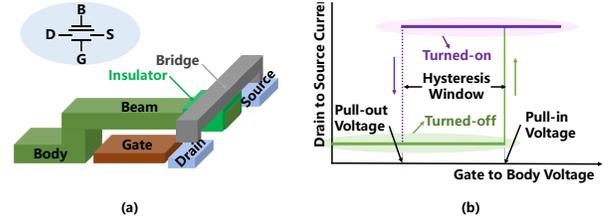

Fig. 3. The 4T NEM relay: (a) Structure; (b) $I_{DS}$–$V_{GB}$ characteristic [12].

## III. PROPOSED NEM-RELAY-BASED 3T2N TCAM

This section proposes the NEM-relay-based 3T2N TCAM, including circuit schematic, write and search operations, refresh operation using one-shot refresh and the overall architecture.

### A. Proposed NEM-Based TCAM Circuit Scheme

Fig. 1 has illustrated the proposed NEM-relay-based 3T2N TCAM circuit. The stored bits (S and $\overline{S}$) input through bitlines (BL and $\overline{BL}$) are written to two parallel NEM relays ($N_1$ and $N_2$) via two write transistors ($T_{W1}$ and $T_{W2}$) controlled by wordline (WL). Because 4T NEM relays have no $V_{th}$ drop when passing a high level voltage, SL and $\overline{SL}$ can be connected with the 4T NEM relay drain directly without voltage reduction. The NEM relay source is connected to the gate of another transistor ($T_S$) to control the matchline (ML) discharging during a mismatch case. In addition to the complementary bits, $N_1$ and $N_2$ can both store OFF-state or logic '0' to represent the '*don't care*' state. Therefore, S/$\overline{SL}$ and $\overline{S}$/SL with $T_S$ work as two pull down paths between ML and the ground. If there exists at least one ON pull down path, mismatch occurs and the voltage of the pre-charged ML settles towards the ground. It is thus clear that the 3T2N TCAM cell provides an XNOR output $\overline{S \oplus SL}$ at ML.

### B. Write Operation

Write operation of the 3T2N TCAM is similar to that of other TCAMs. WL is first driven to VDD, then the write voltages (VDD for '1' and GND for '0') are applied to the bitlines (BL and $\overline{BL}$) to switch the NEM relay state. Other lines are grounded. Note that the charges on the capacitance between gate and body ($C_{GB}$) represent stored state.

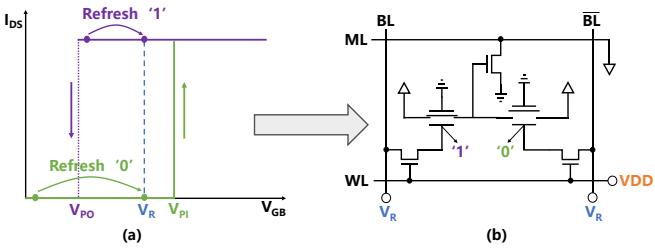

Fig. 4. The one-shot refresh operation: (a) Theory; (b) Application.

### C. Search Operations

Search operation of the 3T2N TCAM is carried out as follows. ML is first pre-charged to VDD, and then the input SL and $\overline{SL}$ are driven to corresponding voltages. If both pull-down paths are OFF, match occurs. Otherwise, if there exists at least one ON pull-down paths, mismatch is triggered and ML is pulled down towards GND. Note that unlike RRAM-based TCAMs, the 3T2N TCAM has negligible leakage paths through between ML and the ground during search operation.

### D. Refresh Operation

The 4T NEM relay is a volatile device, so refresh operation is required to avoid the stored data from being lost. Conventional refresh operation is carried out in a row-by-row manner: first read-out and then write-back. However, it is apparent that this refresh manner not only consumes high energy but also stalls many normal search requests.

Excitingly, a recent work in [14] proposed one-shot refresh (OSR) technology which can significantly reduce the refresh overhead for volatile devices with hysteresis characteristics. As shown in Fig. 4(a), when maintaining the NEM relay ON-state of bit '1', if $C_{GB}$ of NEM relay is not charged to VDD but a certain voltage called refresh voltage ($V_R$) within the hysteresis window, we find that ON-state will not be changed. Similarly, the OFF-state will not be changed when applying $V_R$ to NEM relay gate. In other words, refreshing methods of '0' and '1' can share the same operation, so the entire array could be refreshed simultaneously, and the prior read operation in row-by-row refresh is not needed at all. In this way, the number of refresh operations is reduced from N (the number of rows in the array) to only 1, and the refresh power becomes ultra-low (e.g. 241nW for a 32KB eDRAM array [14]) and almost no normal access will be stalled.

Fig. 4(b) shows the OSR applied to the 3T2N TCAM. Although two NEM relays in one 3T2N TCAM cell store complementary bits, they can still be refreshed in the same voltage configuration.

## IV. 3T2N TCAM BENCHMARKING

This section evaluates the performance of the proposed NEM relay-based 3T2N TCAM. The NEM relay simulation model and other benchmarking settings will be introduced first. The performance of write and search operation, including energy and latency, will then be evaluated and analyzed.

### A. Benchmarking Settings

This paper utilizes the NEM relay SPICE model from [15][16] for circuit analysis. This model has been widely used in prior works [17][18]. The simplified concept of the ON and OFF states is shown in Fig. 5. Many works have illustrated the relationship between these parameters and the device dimensions [12]. Note that NEM relay scales well and [18] has shown how to scale down NEM relays to achieve device-circuit co-design. Here we set the gate thickness as 7.6nm, and the other key simulation parameters are listed in TABLE I.

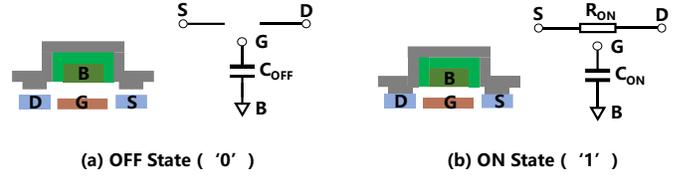

Fig. 5. NEM relay circuit modeling.

TABLE I
NEM RELAY SIMULATION PARAMETER

| Parameter | $V_{PI}$ | $V_{PO}$ | $C_{ON}$ | $C_{OFF}$ | $R_{ON}$ | $\tau_{mech}$ |
|---|---|---|---|---|---|---|
| Value | 0.53V | 0.13V | 20aF | 15aF | 1kΩ | 2ns |

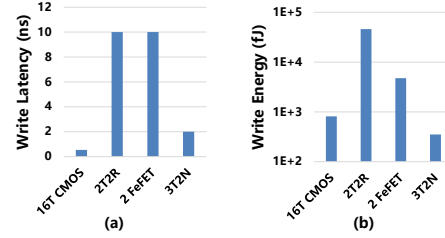

Fig. 6. 3T2N TCAM write performance (for a row): (a) latency; (b) energy.

Note that the mechanical switching latency $\tau_{mech}$ is in the range of nanoseconds. This speed could be higher than some nonvolatile memory devices (~10ns).

The array size is set to 64×64 (4Kb), and SPICE simulation is carried out on it. Each line has been added with a parasitic capacitor scaled by the TCAM cell size. Here we compare the 3T2N TCAM design with 16T SRAM TCAM, 2T2R TCAM, and the 2FeFET TCAM designs for evaluation. All CMOS transistors in these designs are based on the 45nm low-power PTM model with a minimized transistor size for higher density [19]. For the RRAM TCAM, this paper adopts the model and parameters from [8][20], including $R_{ON}/R_{OFF}$ (20kΩ/2MΩ), set/reset voltage (1.8V/1.2V), and write time (10ns). For The FeFET TCAM, the Preisach FeFET model from [11] is utilized and the write voltage and write time is set to ±4V and 10ns, respectively, which is consistent with those in [2][8].

### B. Refresh Scheme

The 3T2N TCAM requires refresh operation to avoid the stored data from being lost, and here we prove that, when using OSR, the refresh overhead can be negligible. $V_R$ is set to 500mV. This value is a little smaller than $V_{PI}$ for noise and variation consideration. With 1.0V supply voltage, one OSR operation consumes about 520fJ for the 3T2N TCAM, which is less than the power consumption of writing 2 rows, as to be shown in the next subsection. Besides, the simulated retention time for the 3T2N TCAM is about 26.5μs. In other words, the refresh power for the 3T2N TCAM array is only about 19.6nW. For most applications, this overhead is reasonably low.

### C. Write Evaluation

The write evaluation is carried out to one row of the array. Fig. 6(a) shows the comparison of write latency between different TCAMs. Note that the write latency is in the array level, and that the time for addressing and decoding (~1ns for 64×64 array) is not added up because it is the same for all TCAMs with the same array size. From the results in Fig. 6(a), it is not surprising that the SRAM TCAM is the fastest one with only ~0.5ns write time. The 3T2N TCAM is the second fastest one with ~2ns write time, close to $\tau_{mech}$. The 2T2R TCAM and the 2FeFET TCAM take longer time, ~10ns, which is much larger than the delay of typical peripherals. Such a significant write time improvement from 10ns to 2ns by the proposed 3T2N TCAM may bring well-observed overall performance enhancement, especially for write-intensive applications.

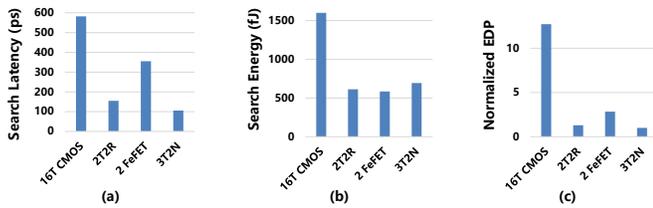

Fig. 7. Search performance evaluation of 3T2N TCAM: (a) Search latency; (b) Search energy; (c) Normalized search EDP.

Fig. 6(b) summarizes the write energy of different TCAMs for writing a row. The 2T2R TCAM requires ~46pJ energy due to RRAM current-driven write mechanism. For the 2-FeFET TCAM, the need of 4V write voltage makes the energy for charging the bitline capacitance high and the energy consumption is ~4.7pJ. The SRAM TCAM operates at only 1V, but the larger cell size brings the largest parasitic capacitance so 0.81pJ is consumed. For the 3T2N NEM relay, thanks to the capacitive write load and the CMOS-compatible 1V operating voltage, only 0.35pJ is consumed for write a row of data. The energy efficiency improvement is 2.31x, 131x, and 13.5x over the SRAM, RRAM, and FeFET-based TCAMs, respectively.

### D. Search Evaluation

Unlike the write operations, VDD during a search operation could be set to 1V for all TCAM designs. The search latency is measured for the worst case, i.e. with only 1-bit mismatch cell discharging ML. Fig. 7(a) shows the search energy evaluation of different TCAMs. It is observed that the search speed of the 3T2N TCAM is 5.50x, 1.47x, and 3.36x faster than that of the SRAM, RRAM, and FeFET-based TCAMs, respectively. This is because the ON-state resistance of NEM relay (~1kΩ) is much smaller than that of MOSFET, leading to less RC discharge time to distinguish matched or mismatched results. RRAM-based TCAM could also be made faster in search operations with lowered resistance, but at the cost of increased write power.

Fig. 7(b) shows the search energy. Results show that the search energy of the 3T2N TCAM is 2.31x, 0.88x, and 0.84x of SRAM, RRAM, and FeFET-based TCAMs, respectively. Note that the 3T2N TCAM needs more search energy than the 2-FeFET TCAM and the 2T2R TCAM because of a larger cell footprint that causes more bitline parasitic capacitance.

Fig. 7(c) compares the search energy-delay-product (EDP) as a more balanced metric. SRAM and FeFET-based TCAMs exhibit 12.7x and 2.83x higher EDPs than the proposed 3T2N TCAM, respectively. RRAM-based TCAM has an EDP equal to 1.3x of the proposed 3T2N TCAM, at the assumption of no device variations. When variations are considered, the settling of the matchline in RRAM-based TCAM will be more difficult to identify between matched and mismatched scenarios, and NEM-relay-based TCAM shows better EDP.

## V. CONCLUSION

This paper has presented a novel dynamic ternary content-address memory (TCAM) using nanoelectromechanical (NEM) relay devices. The proposed TCAM cell structure and the array refreshing scheme harness the unique NEM relay device characteristics, including the nearly zero OFF-state leakage, ultra-low ON-state resistance, and sharp ON-OFF transition with a hysteresis window. The proposed TCAM is capable of achieving low power, high speed and high density. The refresh of the dynamic memory is taken care by the one-shot refresh (OSR) scheme. SPICE simulations have been carried out for benchmarking. Results have shown 2.31x, 131x, and 13.5x lower write energy and 12.7x, 1.30x, and 2.83x less search EDP when comparing with SRAM, RRAM, and FeFET-based TCAMs, respectively.

ACKNOWLEDGMENT

This work is supported in part by The National Key Research and Development Program of China (2019YFA0706100, 2018YFB2202802), NSFC (61874066, 61720106013), Tsinghua University Initiative Scientific Research Program, ICFC, and the National Training Program of Innovation and Entrepreneurship for Undergraduates. H. Zhong and S. Cao contribute equally to this work.

REFERENCES

[1] Y.-J. Chang, "A high-performance and energy-efficient TCAM design for IP-address lookup," IEEE Trans. Circuits Syst. II Express Briefs, vol. 56, no. 6, pp. 479–483, 2009.
[2] K. Ni et al., "Ferroelectric ternary content-addressable memory for one-shot learning," Nat. Electron., vol. 2, no. 11, pp. 521–529, 2019.
[3] K. Pagiamtzis and A. Sheikholeslami, "Content-addressable memory (CAM) circuits and architectures: A tutorial and survey," IEEE J. Solid-State Circuits, vol. 41, no. 3, pp. 712–727, 2006.
[4] V. Vinogradov, et al., "Dynamic ternary CAM for hardware search engine," Electron. Lett., vol. 50, no. 4, pp. 256–258, 2014.
[5] S. Matsunaga, et al., "Design of a nine-transistor/two-magnetic-tunnel-junction-cell-based low-energy nonvolatile ternary content-addressable memory," Jpn. J. Appl. Phys., vol. 51, no. 2S, p. 02BM06, 2012.
[6] J. Li, R. K. Montoye, M. Ishii, and L. Chang, "1 Mb 0.41μm² 2T-2R cell nonvolatile TCAM with two-bit encoding and clocked self-referenced sensing," IEEE J. Solid-State Circuits, vol. 49, no. 4, pp. 896–907, 2013.
[7] X. Yin, M. Niemier, and X. S. Hu, "Design and benchmarking of ferroelectric FET based TCAM," in Design, Automation & Test in Europe Conference & Exhibition (DATE), 2017, 2017, pp. 1444–1449.
[8] X. Yin, K. Ni, D. Reis, S. Datta et al., "An ultra-dense 2FeFET TCAM design based on a multi-domain FeFET model," IEEE Trans. Circuits Syst. II Express Briefs, vol. 66, no. 9, pp. 1577–1581, 2018.
[9] K. Ni, X. Li, J. A. Smith, M. Jerry, and S. Datta, "Write Disturb in Ferroelectric FETs and Its Implication for 1T-FeFET AND Memory Arrays," IEEE Electron Device Lett., vol. 39, no. 11, pp. 1656–1659, 2018.
[10] X. Li, J. Wu, K. Ni, S. George, K. Ma, et al., "Design of 2T/cell and 3T/cell nonvolatile memories with emerging ferroelectric fets," IEEE Des. Test, vol. 36, no. 3, pp. 39–45, 2019.
[11] K. Ni, M. Jerry, J. A. Smith, and S. Datta, "A Circuit Compatible Accurate Compact Model for Ferroelectric-FETs," in 2018 IEEE Symposium on VLSI Technology, 2018, pp. 131–132.
[12] F. Chen et al., "Integrated circuit design with NEM relays," in Proceedings of the 2008 IEEE/ACM ICCAD, 2008, pp. 750–757.
[13] U. Sikder and T.-J. K. Liu, "Design optimization for NEM relays implemented in BEOL layers," in 2017 IEEE SOI-3D-Subthreshold Microelectronics Technology Unified Conference (S3S), 2017, pp. 1–3.
[14] H. Zhong et al., "One-Shot Refresh: A Low-Power Low-Congestion Approach for Dynamic Memories," in IEEE Trans. Circuits Syst. II Express Briefs, vol. 67, no. 12, pp. 3402-3406, 2020.
[15] C. Chen et al., "Efficient FPGAs using nanoelectromechanical relays," in Proceedings of the 18th annual ACM/SIGDA international symposium on Field programmable gate arrays, 2010, pp. 273–282.
[16] S. Chong et al., "Nanoelectromechanical (NEM) relays integrated with CMOS SRAM for improved stability and low leakage," in 2009 International Conference on Computer-Aided Design, 2009, pp. 478–484.
[17] X. Huang et al., "A nanoelectromechanical-switch-based thermal management for 3-D integrated many-core memory-processor system," IEEE Trans. Nanotechnol., vol. 11, no. 3, pp. 588–600, 2012.
[18] H. Zhong, M. Gu, J. Wu, H. Yang, and X. Li, "Design of almost-nonvolatile embedded DRAM using nanoelectromechanical relay devices," in 2020 Design, Automation & Test in Europe Conference & Exhibition (DATE), 2020, pp. 1223–1228.
[19] R. Vattikonda, W. Wang, and Y. Cao, "Modeling and minimization of PMOS NBTI effect for robust nanometer design," in 2006 43rd ACM/IEEE Design Automation Conference, 2006, pp. 1047–1052.
[20] M. A. Lastras-Montano et al, "Architecting energy efficient crossbar-based memristive random-access memories," in Proceedings of the 2015 IEEE/ACM NANOARCH' 15, 2015, pp. 1–6